\documentstyle[amssymb,12pt]{article}

\begin{document}

\title{Black Hole Entropy Associated with Supersymmetric Sigma Model}
\author{M. C. B. Abdalla$^1$, A. A. Bytsenko$^2$ and 
M. E. X. Guimar\~aes$^{1,3}$ \\
\mbox{\small{1. Instituto de F\'{\i}sica Te\'orica/UNESP, S\~ao Paulo, SP, 
Brazil}} \\
\mbox{\small{2. Departamento de F\'{\i}sica, Univ. Estadual de Londrina,
 Londrina, PR, Brazil}} \\
\mbox{\small{3. Departamento de 
Matem\'atica, UnB, Bras\'{\i}lia, DF, Brazil}}}
\date{May, 2003}
\maketitle

\begin{abstract}
By means of an identity that equates elliptic genus partition function of a
supersymmetric sigma model on the $N$--fold symmetric product $S^NX$ of $X$
\,( $S^NX=X^N/S_N$, $S_N$ is the symmetric group of $N$ elements) to the
partition function of a second quantized string theory, we derive the
asymptotic expansion of the partition function as well as the asymptotic for
the degeneracy of spectrum in string theory. The asymptotic expansion for
the state counting reproduces the logarithmic correction to the black hole
entropy.
\end{abstract}




\section{Introduction}

In the correspondence between a black hole and a highly excited string the
black hole horizon is governed by some conformal operator algebra on a
two--dimensional surface. This provides a string representation of black
hole quantum states. Conversely, it may be possible to give a black hole
interpretation of strings \cite{Hooft}. The black hole entropy by counting
the number of excited strings states (statistical interpretation of the
black hole entropy) has been subsequently presented in several papers 
\cite{Bytsenko1,Bytsenko2,Halyo,Halyo1,Horowitz,Bytsenko3,Damour,
Brevik1,Brevik2,Kaul,Rama}. The comparison of the asymptotic state density
of (twisted) $p-$branes and mass level state density of black holes has also
been established in Refs. \cite
{Bytsenko1,Bytsenko4,Bytsenko5,Bytsenko6,Abdalla}. In this work, we
calculate the black hole entropy for a supersymmetric sigma model. In
section 1.1, we set the relevant mathematical method used in this paper.
Then, in section 2 and in section 3, we derive the asymptotic state density
and the black hole entropy, respectively. We end up with some concluding
remarks in section 4.

\subsection{The Mathematical Notation}

We start by considering a supersymmetric sigma model on the $N-$fold
symmetric product $S^NX$ of a K\"{a}hler manifold $X$, which is the orbifold
space $S^N X = X^N/S_N$. $S_N$ is the symmetric group of $N$ elements. The
Hilbert space of an orbifold field theory can be decomposed into twisted
sectors ${\cal H}_{\gamma}$, that are labelled by the conjugacy classes 
$\{\gamma\} $ of the orbifold group $S_N$ \cite{Dixon1,Dixon2,Dijkgraaf1}.
For a given twisted sector one can keep the states invariant under the
centralizer subgroup $\Gamma_{\gamma}$ related to the element $\gamma$. 
Let ${\cal H}_{\gamma}^{\Gamma_{\gamma}}$ be an invariant subspace associated 
with $\Gamma_{\gamma}$; the total orbifold Hilbert space takes the form 
${\cal H}(S^N X)=\bigoplus_{\{\gamma\}} {\cal H}_{\gamma}^{\Gamma_{\gamma}}$. 
Taking into account the group $S_N$ one can compute the conjugacy classes 
$\{\gamma\}$ by using a set of partitions $\{N_n\}$ of $N$, namely 
$\sum_nnN_n=N$, where $N_n$ is the multiplicity of the cyclic permutation 
$(n)$ of $n$ elements in
the decomposition of $\gamma$: $\{\gamma\}=\sum_{j=1}^s(j)^{N_j}$. For this
conjugacy class the centralizer subgroup of a permutation $\gamma$ is 
$\Gamma_{\gamma}=S_{N_1}\bigotimes_{j=2}^s\left(S_{N_j}>\!\!\!\lhd {Z}
_j^{N_j}\right)$ \cite{Dijkgraaf1}, where each subfactor $S_{N_n}$ and 
${Z}_n$ permutes the $N_n$ cycles $(n)$ and acts within one cycle $(n)$
correspondingly. Following the lines of Ref. \cite{Dijkgraaf1} we may
decompose each twisted sector ${\cal H}_{\gamma}^{\Gamma_{\gamma}}$ into 
a product over the subfactors $(n)$ of $N_n-$fold symmetric tensor products, 
${\cal H}_{\gamma}^{\Gamma_{\gamma}}= 
\bigotimes_{n>0}S^{N_n}{\cal H}_{(n)}^{Z_{n}}$,
where $S^N{\cal H}\equiv (\bigotimes^N{\cal H})^{S_N}$.

Let $\chi({};q,y)$ be the partition function for every (sub) Hilbert space
of a supersymmetric sigma model. It has been shown 
\cite{Schellekens1,Schellekens2,Witten,Landweber,Eguchi,Kawai} that the 
partition function coincides with the elliptic genus. If 
$\chi({\cal H}_{(n)}^{{Z}_n};q,y)$
admits the extension 
$\chi({\cal H};q,y)=\sum_{m\geq 0,\ell}C(nm,\ell)q^my^{\ell}$, 
the following result holds (see Refs. \cite{Dijkgraaf1,Dijkgraaf2}): 
\begin{equation}
\sum_{N\geq 0}p^N\chi(S^N{\cal H}_{(n)}^{{\bf Z}_n};q,y)= \prod_{m\geq 0,\ell}
\left(1-pq^my^{\ell}\right)^{-C(nm,\ell)} 
\mbox{,}
\end{equation}
\begin{equation}
W(p; q, y) = \sum_{N\geq 0}p^N\chi(S^NX;q,y)=\prod_{n>0,m\geq 0,\ell}
\left(1-p^nq^my^{\ell}\right)^{-C(nm,\ell)} 
\mbox{,}
\end{equation}
$p={\bf e}[\rho]$, $q ={\bf e}[\tau]$, $y={\bf e}[z]$, and ${\bf e}[x]
\equiv \exp [2\pi i x]$. Here $\rho$ and $\tau$ determine the complexified
K\"{a}hler form and complex structure modulos of ${\bf T}^2$ respectively,
and $z$ parametries the $U(1)$ bundle on ${\bf T}^2$. The Narain duality
group $SO(3,2,{\bf Z})$ is isomorphic to the Siegel modular group $Sp(4, 
{\bf Z})$ and it is convenient to combine the parameters $\rho, \tau$ and a
Wilson line modules $z$ into a $2\times 2$ matrix belonging to the Siegel
upper half--plane of genus two, $\Xi = \left( 
\begin{array}{ll}
\rho\,\,\,z &  \\ 
z\,\,\, \tau & 
\end{array}
\right)$, with $\Im \rho >0,\, \Im \tau >0$, ${\rm det}\,\Im \Xi >0$. The
group $Sp(4, {\bf Z}) \cong SO(3,2,{\bf Z})$ acts on the matrix $\Xi$ by
fractional linear transformations, namely 
$\Xi \rightarrow (A\Xi +B)(C\Xi + D)^{-1}$.

Finally we go into some facts related to orbifoldized elliptic genus of $N=2$
superconformal field theory. The contribution of the untwisted sector to the
orbifoldized elliptic genus is the function $\chi(X;\tau,z)\equiv\phi(%
\tau,z)\equiv \ {\scriptstyle 0}\hskip 1mm \mathop{\fbox{%
\rule[.3cm]{.4cm}{0cm}$\frac{}{}$}}\limits_{\lower 1mm 
\hbox{$\scriptstyle
0$}} \hskip 1mm(\tau,z)$, whereas

\begin{equation}
\phi\left(\frac{a\tau+b}{c\tau+d},\frac{z}{c\tau+d}\right)= \ {\scriptstyle 0
}\hskip 1mm \mathop{\fbox{\rule[.3cm]{.4cm}{0cm}$\frac{}{}$}}\limits_{\lower 
1mm \hbox{$\scriptstyle 0$}} \hskip 1mm(\tau,z){\bf e}\left[\frac{rcz^2}{
c\tau+d}\right], \,\,\,\left( 
\begin{array}{ll}
a\,\,\,b &  \\ 
c\,\,\,d & 
\end{array}
\right)\in SL(2,{\bf Z}) \mbox{,}
\end{equation}
$r=d/2$. The contribution of the twisted $\mu-$sector projected by $\nu$ is 
\cite{Kawai}:

\begin{equation}
\ {\scriptstyle \nu}\hskip 1mm \mathop{\fbox{\rule[.3cm]{.4cm}{0cm}$%
\frac{}{}$}}\limits_{\lower 1mm \hbox{$\scriptstyle \mu$}} \hskip 
1mm(\tau,z)=\phi(\tau,z+\mu\tau+\nu){\bf e} \left[\frac{d}{2}
(\mu\nu+\mu^2\tau+2\mu z)\right] ,\,\,\,\,\,\,\,\mu,\nu\in {\bf Z} \mbox{.}
\label{trans}
\end{equation}
The orbifoldized elliptic genus can be defined by 
\begin{equation}
\phi(\tau,z)_{{\rm orb}}\stackrel{def}{=}\frac{1}{h}\sum_{\mu,
\nu=0}^{h-1}(-1)^ {P(\mu+\nu+\mu\nu)}\ {\scriptstyle \nu}\hskip 1mm 
\mathop{\fbox{\rule[.3cm]{.4cm}{0cm}$\frac{}{}$}}\limits_{\lower 1mm 
\hbox{$\scriptstyle \mu$}} \hskip 1mm(\tau,z) \mbox{,}
\end{equation}
where $P, h$ are some integers.

\section{The Asymptotic Density of States}

If $y={\bf e}[z] =1 $ then the elliptic genus degenerates to the Euler
number or Witten index \cite{Hirzebruch2,Vafa}. For the symmetric product
this gives the following identity 
\begin{equation}
W(p) = \sum_{N\geq 0}p^N\chi(S^NX)=\prod_{n>0} \left(1-p^n\right)^{-\chi(X)} 
\mbox{.}  \label{character}
\end{equation}
Thus this character is almost a modular form of weight $-\chi(X)/2$. Eq. 
(\ref{character}) is similar to the denominator formula of a (generalized)
Kac--Moody algebra \cite{Borcherds,Harvey}. A denominator formula can be
written as follows: 
\begin{equation}
\sum_{\sigma\in {\cal W}}\left({\rm sgn}(\sigma)\right)e^{\sigma(v)}=
e^{v}\prod_{r>0}\left(1-e^{r}\right)^{{\rm mult}(r)} 
\mbox{,}  
\label{denom}
\end{equation}
where $v$ is the Weyl vector, the sum on the left hand side is over all
elements of the Weyl group ${\cal W}$, the product on the right side runs
over all positive roots (one has the usual notation of root spaces, positive
roots, simple roots and Weyl group, associated with Kac--Moody algebra) and
each term is weighted by the root multiplicity ${\rm mult}(r)$. For the $%
su(2)$ level, for example, an affine Lie algebra (\ref{denom}) is just the
Jacobi triple product identity. For generalized Kac--Moody algebras there is
the following denominator formula: 
\begin{equation}
\sum_{\sigma\in {\cal W}}\left({\rm sgn}(\sigma)\right)\sigma\left(e^{v}
\sum_{r} \varepsilon(r)e^{r}\right)=e^{v}\prod_{r>0}
\left(1-e^{r} \right)^{{\rm mult}(r)} 
\mbox{,}
\end{equation}
where the correction factor on the left hand side involves $\varepsilon(r)$
which is $(-1)^n$ if $r$ is the sum of $n$ distinct pairwise orthogonal
imaginary roots and zero otherwise.

The logarithm of the partition function $W(p;q,y)$ is the one--loop free
energy $F(p;q,y)$ for a string on ${\bf T}^2\times X$: 
\begin{equation}
F(p;q,y)=\mbox{log}W(p;q,y)=-\sum_{n>0,m,\ell}C(nm,\ell)\mbox{log}
\left(1-p^nq^my^{\ell}\right)
\end{equation}
\begin{equation}
=\sum_{n>0,m,\ell,k>0}\frac{1}{k}C(nm,\ell)p^{kn}q^{km}y^{k\ell}
=\sum_{N>0}p^N\sum_{kn=N}\frac{1}{k}
\sum_{m,\ell}C(nm,\ell)q^{km}y^{k\ell} 
\mbox{.}
\end{equation}
The free energy can be written as a sum of Hecke operators $T_N$ \cite{Lang}
acting on the elliptic genus of $X$ 
\cite{Borcherds,Gritsenko,Dijkgraaf1}: 
$
F(p;q,y)=\sum_{N>0}p^NT_N\chi(X;q,y)
$. 
\newline
\newline
The goal now is to calculate an asymptotic expansion of the elliptic genus 
$\chi(S^NX;q,y)$. The degeneracies for the sigma model are given by the
Laurent inversion formula: 
\begin{equation}
\chi(S^NX;q,y)=\frac{1}{2\pi i}\oint\frac{W(p,q,y)}{p^{N+1}}dp \mbox{,}
\end{equation}
where the contour integral is taken on a small circle around the origin. Let
the Dirichlet series 
\begin{equation}
{\cal D}(s;\tau,z)=\sum_{(n,m,\ell)>0}\sum_{k=1}^{\infty}\frac{{\bf e} [\tau
mk+z\ell k]C(nm,\ell)}{n^sk^{s+1}} \mbox{}  \label{Dir}
\end{equation}
converge for $0<\Re\,s<\alpha$. We assume that series (\ref{Dir}) can be
analytically continued in the region $\Re\,s\geq-C_0\,\,(0<C_0<1)$ where it
is analytic excepting a pole of order one at $s=0$ and $s=\alpha$, with
residue $\mbox{Res}[{\cal D}(0;\tau,z)]$ and $\mbox{Res}[{\cal D}%
(\alpha;\tau,z)]$ respectively. Besides, let 
${\cal D}(s;\tau,z)={\cal O}(|\Im\,s|^{C_1})$ uniformly in 
$\Re\,s\geq-C_0$ as $|\Im\,s|\rightarrow\infty $,
where $C_1$ is a fixed positive real number. The Mellin--Barnes
representation of the function $F(t;\tau,z)$ has the form 
\begin{equation}
{\cal M}[F](t;\tau,z)=\frac{1}{2\pi i}\int_{\Re\,s=1+\alpha}t^{-s}\Gamma(s) 
{\cal D}(s;\tau,z)ds \mbox{.}  \label{Mellin}
\end{equation}
The integrand in Eq. (\ref{Mellin}) has a first order pole at $s=\alpha$ and
a second order pole at $s=0$. Shifting the vertical contour from $%
\Re\,s=1+\alpha$ to $\Re\,s=-C_0$ (this procedure is permissible) and making
use of the residues theorem one obtains

\[
F(t;\tau,z)=t^{-\alpha}\Gamma(\alpha){\rm Res}[{\cal D}(\alpha;\tau,z)]
+\lim_{s\rightarrow 0}\frac{d}{ds}[s{\cal D}(s;\tau,z)] 
\]
\begin{equation}
-(\gamma+{\rm log}\,t){\rm Res}[{\cal D}(0;\tau,z)] +\frac{1}{2\pi i}%
\int_{\Re\,s=-C_0}t^{-s}\Gamma(s){\cal D}(s;\tau,z)ds 
\mbox{,}  
\label{F}
\end{equation}
where $t\equiv 2\pi(\Im\rho-i\Re\rho)$. The absolute value of the integral
in (\ref{F}) can be estimated to behave as 
${\cal O}\left((2\pi\Im\,\rho)^{C_0}\right)$. We are ready now to state the 
main result. \newline

In the half--plane $\Re t>0$ there exists an asymptotic expansion for $%
W(t;\tau,z)$ uniformly in $|\Re\rho|$ for $|\Im\rho|\rightarrow 0, \,\, |%
\mbox{arg}(2\pi i\rho)|\leq \pi/4,\,\,|\Re\rho|\leq 1/2$ and given by 
\[
W(t;\tau,z)={\bf e}\left[\frac{1}{2\pi i}\left\{ {\rm Res}[{\cal D}
(\alpha;\tau,z)]\Gamma(\alpha)t^{-\alpha} -{\rm Res}[{\cal D}(0;\tau,z)]{\rm 
{log}}t \right.\right. 
\]
\begin{equation}
\left.\left. -\gamma{\rm Res}[{\cal D}(0;\tau,z)] +\lim_{s\rightarrow 0}
\frac{d}{ds}[s{\cal D}(s;\tau,z)]+
{\cal O}\left( |2\pi\Im\tau|^{C_0}\right)\right\}
\right] 
\mbox{.}
\end{equation}

The asymptotic expansion at $N \rightarrow \infty$ for the elliptic genus
(see also Refs. \cite{Bytsenko7,Bytsenko3} is given by the following formula

\[
\chi(S^NX;\tau,z)_{N\rightarrow \infty}={\cal C}(\alpha;\tau,z) N^{(2{\rm Res} 
[{\cal D}(0;\tau,z)]-2-\alpha)/(2(1+\alpha))} 
\]
\begin{equation}
\times {\bf e}\left[\frac{1+\alpha}{2\pi i\alpha}\left({\rm Res} [{\cal D}
(\alpha;\tau,z)]\Gamma(1+\alpha)\right)^ {1/(1+\alpha)}N^{\alpha/(1+\alpha)}
\right]\left[1+{\cal O}(N^{-k})\right] 
\mbox{,}  
\label{M1}
\end{equation}
\[
{\cal C}(\alpha;\tau,z)=\left\{{\rm Res}[{\cal D}(\alpha;\tau,z)]
\Gamma(1+\alpha)\right\} ^{(1-2{\rm Res}[{\cal D}(0;q,y)])/(2+2\alpha))} 
\]
\begin{equation}
\times {\bf e}\left[\frac{1}{2\pi i}\left(\lim_{s\rightarrow 0}\frac{d}{ds} 
[s{\cal D}(0;\tau,z)]-\gamma{\rm Res}[{\cal D}(0;\tau,z)]\right)\right] 
\left[2\pi(1+\alpha)\right]^{1/2} 
\mbox{,}  
\label{M2}
\end{equation}
where $k<\alpha/(1+\alpha)$ is a positive constant. In the above formulae
the complete form of the prefactor ${\cal C}(\alpha;\tau,z)$ appears. 
The results (\ref{M1}), (\ref{M2}) have an universal character for all 
elliptic genera associated to Calabi--Yau manifolds.

\section{The Black Hole Entropy}

In the context of string dynamics the asymptotic state density gives a
precise computation of the free energy and entropy of a black hole. The
corresponding black hole entropy ${\cal S}(N)$ takes the form: 
\begin{equation}
{\cal S}(N) = {\rm log}\, \chi(S^NX;\tau,z) \simeq  
{\cal S}_0 + {\cal A}(\alpha){\rm log} ({\cal S}_0) + ({\rm Const.}) 
\mbox{,}
\label{entropy}
\end{equation}
\begin{equation}
{\cal A}(\alpha) = (2\alpha)^{-1}(2{\rm Res}[{\cal D}(0;\tau,z)]- 2-\alpha) 
\mbox{.}
\end{equation}
The leading term in Eq. (\ref{entropy}) is 
${\cal S}_0 = {\cal B}(\alpha)N^{\delta(\alpha)}$, where 
\begin{equation}
{\cal B}(\alpha) = \frac{1}{\delta (\alpha)}\left({\rm Res}[{\cal D}
(\alpha;\tau,z)]
\Gamma(1+\alpha)\right)^{\delta(\alpha)/{\alpha}}\,,\,\,\,\,\,\,\,\,\, 
\delta(\alpha) = \frac{\alpha}{1+\alpha} 
\mbox{,}
\end{equation}
while ${\cal A}(\alpha)$ is the coefficient of the logarithmic correction to the
entropy.

The asymptotic state density at level $N$ ($N \gg 1$) for fundamental $p-$
branes compactified on manifold with topology ${\bf T}^p\times {\bf R}^{d-p}$
can be calculate within the semiclassical quantization scheme (see for
details Refs. \cite{Bytsenko1,Bytsenko7}). The coefficient ${\cal A}(p)$ in this
case takes the form: 
\begin{equation}
{\cal A}(p) = (2p)^{-1}\left( Z_p (0) - 2-p\right) 
\mbox{,}
\end{equation}
where $Z_p(s)$ is the $p-$dimensional Epstein zeta function. Since 
$Z_p(s=0)= -1$ we have ${\cal A}(p)= -(d+1)/(2p)$. In string theory, in the
case of zero modes, the dependence on embedding spacetime can be eliminate 
\cite{Rama}. In fact, the coefficient logarithmic correction ${\cal A}(p)$ 
becomes $-3/2$, which agrees with the results obtained in the spin network
formalism. The coefficient of the logarithmic correction to the
supersymmetric string entropy, ${\cal A}(\alpha)$, depends on the complex
dimension $d$ of a K\"{a}hler manifold $X$.

Using the transformation properties (\ref{trans}) in Eqs. (\ref{M1}), (\ref
{M2}) one can obtain the asymptotic expansion for the orbifoldized state
density. Thus starting with the expansion of the state density of the
untwisted sector we can compute the asymptotics of the state density of the
twisted sector.

\section{Concluding Remarks}

Our results can be used in the context of the brane method's calculation of
the ground state degeneracy of systems with quantum numbers of certain BPS
extreme black holes \cite{Callan,Maldacena, Vafa2,Halyo}. We note here the
BPS black hole in toroidally compactified $(M={\bf T}^{5}\times X^{5})$ type 
II string theory. One can construct a brane configuration such that the
corresponding supergravity solutions describe five--dimensional black holes.
Five branes and one brane are wrapped on ${\bf T}^{5}$ and the system is
given by the Kaluza--Klein momentum $N$ in one of the directions. Thus black
holes in these theories can carry both an electric charge $Q_{F}$and
an axion charge $Q_{H}$. The brane picture gives the entropy in
terms of partition function $W(t)$ for a gas of $Q_{F}Q_{H}$ species of
massless quanta: 
$W(t)=\prod_{{\bf n}\in {\bf Z}^{m}/\{{\bf 0}\}}\left[
1-\exp \left( -t\omega _{{\bf n}}({\bf a},{\bf g})\right)\right] 
^{-({\rm dim}\,M - m-1)}
$,
where $t=y+2\pi ix$, $\Re\, t>0$,\,\,
$
\omega _{{\bf n}}({\bf a},{\bf g})=\left( \sum_{j}a_{j}(n_{j}+
{\rm g}_{j})^{2}\right)^{1/2}
$,\,\, 
${\rm g}_{j}$ and $a_{j}$ are some real numbers. 
For unitary conformal theories of fixed central charge $c$
Eq. (\ref{M1}) represents the degeneracy of the state $\chi (N)$ with
momentum $N$ and for $N\rightarrow \infty $ one has \cite{Bytsenko5}: 
\begin{equation}
{\rm log}\chi (N) \simeq 2\sqrt{\Lambda \zeta _{R}(2)cN}
-\frac{\Lambda c +3}{4}
{\rm log}(N) 
\mbox{,}
\end{equation}
where $\Lambda = ({\rm dim}\,M-m-1)/4$ and $\zeta_{R}(s)$ is the 
Riemann zeta function. 
The entropy takes the form
\begin{equation}
{\cal S}(N)={\rm log} \chi (N)\simeq {\cal S}_{0} +
{\cal A}{\rm log} ({\cal S}_{0}) 
\mbox{,}
\end{equation}
where for $\Lambda =1$ we have
\begin{equation}
{\cal S}_{0}=2\pi \sqrt{cN/6}\,,\,\,\,\,\,\,\,\,\,
{\cal A} = -\frac{c+3}{2}
\mbox{.}
\end{equation}
Following Ref. [36], we can put $c=3Q_{F}^{2}+6,\,\, N=Q_{H}$ and get the
growth of the elliptic genus (or the degeneracy of BPS solitons) 
for $N=Q_{H}\gg 1$. However, this result is incorrect when the black hole
becomes massive enough for its Schwarzschild radius to exceed any
microscopic scale such as the compactification radii \cite{Maldacena,Halyo}.
Such models, stemming from string theory, would therefore be incompatible;
in view of the present result, this might be presented as a usefull
constraint for the underlying microscopic field theory.

Finally, note that for a Calabi--Yau space the $\chi_y-$genus 
\cite{Hirzebruch1} is a weak Jacobi form of weight zero and index $d/2$ and it
transforms as $\chi_y(T_X)=(-1)^{r-d}y^r\chi_{y^{-1}}(T_X)$. This relation
can also be derived from the Serre duality $H^j(X;{\bigwedge}^sT_X)\cong
H^{d-j}(X;{\bigwedge}^{r-s}T_X)$. For $q=0$ the elliptic genus reduces to a
weighted sum over the Hodge numbers, namely $\chi(X;0,y)=%
\sum_{j,k}(-1)^{j+k}y^{j-\frac{d}{2}}h^{j,k}(X)$. For the trivial line
bundle the symmetric product (\ref{character}) can be associated with the
simple partition function of a second quantized string theory.

\section*{Acknowledgements:}

We would like to thank referee for Ref. \cite{Vafa2} and for
constructive comments on BPS
black holes in toroidally compactified type II string theory.
A. A. B. and M. E. X. G. would like to thank CAPES for partial financial
support in the context of the PROCAD programme. M. E. X. G. would like to
thank FAPESP for partial financial support.

\end{document}